\documentclass[preprint,showpacs,preprintnumbers,amsmath,amssymb]{revtex4}

\usepackage{graphicx}% Include figure files
\usepackage{dcolumn}% Align table columns on decimal point
\usepackage{bm}% bold math

\begin{document}

%\preprint{APS}

\title{The Harmonic Measure for critical Potts clusters}

%\author{David A. Adams$^{1}$, Yen Ting Lin$^{1}$, Leonard M. Sander$^{1,2}$, Robert M. Ziff$^{2,3}$}

\date{\today}% It is always \today, today,
\begin{abstract}
We present a  technique, which we call ``etching," which we use to study the harmonic measure of Fortuin-Kasteleyn  clusters in the $Q$-state Potts model for $Q=1-4$. The harmonic measure is the probability distribution of random walkers diffusing onto the perimeter of a cluster.  We use etching to study regions of clusters which are extremely unlikely to be hit by random walkers, having hitting probabilities down to $10^{-4600}$.  We find good agreement between the theoretical predictions of Duplantier and our numerical results for the generalized dimension $D(q)$, including regions of small and negative $q$.  
\end{abstract}

\author{D. A. Adams}
 \affiliation{Department of Physics, University of Michigan, Ann Arbor MI 48109-1040}
\author{Yen Ting Lin}
\affiliation{Department of Physics, University of Michigan, Ann Arbor MI 48109-1040}
\author{L. M. Sander}
 \altaffiliation[ Also at ]{Michigan Center for Theoretical Physics, University of Michigan, Ann Arbor MI 48109-1040}  %  optional
\affiliation{Department of Physics, University of Michigan, Ann Arbor MI 48109-1040}
\author{R. M. Ziff}
 \altaffiliation[ Also at ]{Michigan Center for Theoretical Physics, University of Michigan, Ann Arbor MI 48109-1040}  %  optional
\affiliation{Department of Chemical Engineering, University of Michigan, Ann Arbor MI 49109-2136}

%\address{$^{1}$Department of Physics, University of Michigan, Ann Arbor MI 48109-1040 \\
%$^{2}$Michigan Center for Theoretical Physics, University of Michigan, Ann Arbor MI 48109-1040
%\\$^{3}$Department of Chemical Engineering, University of Michigan, Ann Arbor MI 49109-2136}

\maketitle

\section{Introduction}

\subsection{Potts model}
The $Q$-state Potts model, a generalization of the Ising model %\cite{Ising25} 
to $Q$ different spins, has been the subject of considerable interest \cite{Wu82}.  Two important cases are $Q=1$ and $Q=2$, which correspond to percolation \cite{StaufferBook94} and the Ising model, respectively. When a Potts  system is prepared at its critical temperature, subsets of the clusters of like spins, the Fortuin-Kastelyn (FK) clusters \cite{Fortuin72, Kasteleyn69} (to be defined below), are self-similar fractals \cite{Mandelbrot82}. For  $Q=1$ the FK clusters are the same as the usual percolation clusters. In this paper, we will study the harmonic measure of the hulls of these fractal clusters for $Q=1,2,3,4$. 

The harmonic measure may be thought of as the distribution of the surface electric field on a charged conductor.  Since the Laplace equation and the steady-state diffusion equation are identical in form, the harmonic measure is also equal to the distribution of probabilities of random walkers diffusing far from the cluster onto a given section of the hull. In this article, we use a biased random-walk sampling technique to obtain the harmonic measure. We also review other methods for measuring small probabilities and give details of our algorithms. 

The harmonic measure is of practical interest because of its relation to the anomalous frequency dependence of the impedance of rough electrodes \cite{Halsey92} and because of its obvious connection to processes that involve absorption of diffusing particles such as  catalysis \cite{Sapoval01}.  It has a deep connection to the structure of diffusion-limited aggregates (DLA) \cite{Witten81}, since the harmonic measure determines where each walker will land; that is, for DLA it is the growth probability. In the case of critical Potts clusters and DLA, the harmonic measure is multifractal \cite{Mandelbrot90}. Advances in conformal field theory and  Schramm-Loewner evolution (SLE) have brought about renewed interest in the harmonic measure. In particular, certain aspects of the harmonic measure for Potts clusters can be computed  in the continuum limit using these methods \cite{Duplantier99, Duplantier00, Belikov08, Duplantier08, Bettelhiem05, Gruzberg06}.

Numerical investigation of  the harmonic measure of percolation \cite{Meakin86} and DLA \cite{Meakin86, Ball90, Hanan08} clusters is difficult because the measure has a huge dynamic range for  systems of even moderate size. In references \cite{Meakin86, Ball90, Hanan08}  one of two methods were used: The first is the straightforward one of releasing a large number of random walkers far from the cluster and determining where they land. The second uses relaxation or equivalent algorithms to solve the Laplace equation. The random walker method can only measure probabilities down to about $10^{-10}$ and samples a very small part of the measure for clusters of reasonable size. Relaxation-like methods are computationally costly and limited to small clusters, and give similar lower limits on the probabilities that can be measured.

For DLA  it is possible to go to much smaller probabilities by using the method of iterated conformal maps \cite{Hastings98, Davidovitch00, Davidovitch01}. However, this technique is only capable of treating moderate size  clusters \cite{Jensen02}.  In an earlier publication we generalized the random walker method and gave a  technique capable of obtaining probabilities down to $10^{-300}$ for any fractal. We applied it to FK clusters for percolation and the Ising model  \cite{Adams08}. This paper describes a further development of those techniques.

\begin{figure}
\includegraphics[width=0.48\textwidth]{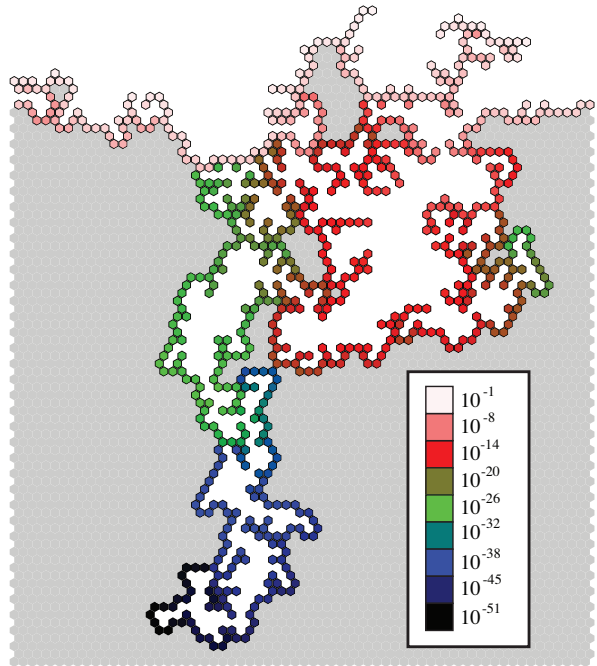}
\caption{\label{fig:picture}  (Color online) The harmonic measure for the complete perimeter of a small, $W=64$, percolation cluster. The solid grey regions represents the area that is inaccessible to the random walkers diffusing from above the cluster. Every perimeter site is colored according to its measure. The computation was performed using the etching method described below. Note that the measure on this small cluster spans $50$ orders in magnitude.}
\end{figure}

\subsection{Generalized dimensions}
The harmonic measure, the distribution of probabilities that random walkers will hit a given site on the perimeter of a cluster, is very complicated and varies wildly for the cases we are studying; see Figure \ref{fig:picture}. A popular and useful way to characterize it is  in terms of the generalized dimension, $D(q)$,  of the measure. We define these objects as follows: We cover the hull with boxes of length $L$. With each box we  associate a probability, $p_i$, which is the sum of the measure over the sites within the box. We then define  a function $Z_L(q)$, sometimes called the partition function:

\begin{equation}
\label{eq:Zq}
Z_L(q) = \sum_i p_{i}^q
\end{equation}
where $q$ is some power  \cite{Halsey86}.  If the object in question is fractal, then the partition function will follow a power-law in $L$:
\begin{equation}
\label{eq:Zq2}
Z_L(q) \sim (R/L)^{-\tau_q} = (R/L)^{-(q-1)D(q)}
\end{equation}
for $(R/L) \rightarrow \infty$, where $R$ is the size of the cluster. For integer $q$, $D(q)$ corresponds to the fractal dimension of the $q$-point correlation function. There are special values of $D(q)$. $D(0)$ is the fractal, box-counting, dimension of the hull. Also $D(1)$, the information dimension, is always unity by Makarov's theorem \cite{Makarov85}.  A related function is the singularity spectrum $f(\alpha)$,  the Legendre transform of $D(q)$:

\begin{equation}
\label{f_alpha}
f(\alpha) = q \frac{d \tau}{d q} - \tau, \quad \alpha = \frac{d \tau}{d q}.
\end{equation}
In this article, we will focus exclusively on the $D(q)$. The singularity spectrum can be derived from our results using Eq.~(\ref{f_alpha}).

\section{Models}

\subsection{Simulations of FK clusters for the Potts model}

We produce critical Potts clusters in two ways. For percolation, we use the Leath algorithm \cite{Leath76}. The algorithm starts with a single active site; we attempt to turn its neighbors into active sites with probability $p$. If a conversion attempt fails, the site is labeled inactive. The process is repeated with neighbors of the active sites which have not been labeled as inactive. The process continues until there are no new active sites. If $p$ equals $p_c$, the percolation threshold, a critical percolation cluster is produced. The outer layer of active sites is called the complete perimeter.  Its fractal dimension is denoted $D_H$. The cluster of active sites is surrounded by a single layer of inactive sites; this layer is called the accessible (or exterior) perimeter \cite{Grossman87}, and has a fractal dimension denoted $D_{EP}$. The accessible perimeter of interest because, unlike the complete perimeter, it is expected to have a well-behaved limit when clusters are very large and are rescaled. The harmonic measure has been determined in this limit for the accessible perimeter of Potts clusters \cite{Duplantier99,Duplantier00}.  

To obtain critical Potts clusters for $Q=2,3,4$, we grow equilibrated FK clusters using the Swendsen-Wang (SW) algorithm \cite{Swendsen87}.  For any configuration of spins, FK clusters are subsets of clusters of like spins formed by a  bond percolation process. That is, we consider the clusters formed when adjacent spins are connected with probability $p_c(Q) = 1-\exp[-K_c(Q)]$ where $K_c(Q)$ is the critical coupling constant. For $Q=2,3,4$, on the triangular lattice, $p_c(Q)$ is known to be $1~-~1/\sqrt3$, $1- 1/[1+\frac{1}{2}\sqrt{3}\sec({\pi}/{18})]$, and $1/2$, respectively \cite{kim1974exact}. To obtain the equilibrium ensemble of FK clusters we iterate two steps until the system settles down (see below).  The first step   takes every current FK cluster  and replaces the spin  with one of the $Q$ possible values, at random.  In the second step, the bonds connecting the clusters are discarded and bond percolation is performed again, with $p=p_c(Q)$, on all neighboring sites with the same spin.   The process is then repeated by updating the spins on the newly formed clusters. These two steps together constitute  a spin update.

%Our rare event methods can only be applied to site percolation like problems, so we must turn the bond clusters into site clusters. We do this by creating a lattice with twice the density of the original lattice. We add sites on the center of the bonds and where two bonds meet. This successfully turns the bond clusters into site clusters which preserves the fractal boundary. 

\subsection{Parameters and Observables}

\begin{figure}[t]
\includegraphics[width=0.48\textwidth]{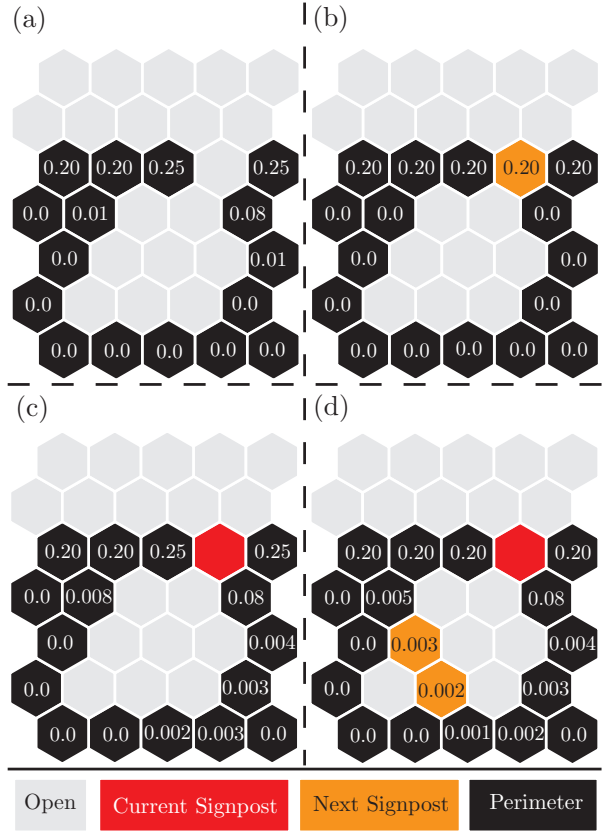}
\caption{\label{fig:Signpost} (Color online) The signpost method. The system is periodic in the horizontal direction. (a) First (probe) step: $N$ random walkers are released from the top row and absorb onto the perimeter sites.  (b) We choose the first probability threshold as $0.1$. Using this threshold, we connect the the bounding sites using signpost sites. In the second (measurement) step we send $N$ more random walkers from above which can absorb onto the signpost or perimeter sites. (c) Second probe step: the random walkers  are launched from the signpost sites in the previous measurement step.  The weights of the walkers released in this step have a weight of $p/N$ where $p$ is the fraction of the random walkers that hit the signpost site in  (b). (d) The second threshold, $0.01$, is used to determine the location of the new signpost sites.  }
\end{figure}

We grew critical Potts clusters for $Q=1-4$ on the triangular lattice, as described above. We chose to use a triangular lattice rather than  a square lattice because the square lattice does not allow diffusion into fjords bounded by diagonal entrances. We use the width of the system, $W$, as the characteristic length.  The clusters we want span in the width direction but not in the height direction. To make sure the clusters will only span in one direction, we chose very large aspect ratios. The height of the lattices were $100 W$ and $8 W$ for $Q=1$ and $Q>1$, respectively.  We looked at six different system widths, $W= 128, 256, 512, 1024, 2048, 4096$. Because FK clusters are intrinsically bond clusters, we needed to use a trick to turn them into site clusters.  We created a lattice twice as dense as the original and marked every site at the center of a bond and every site where two bonds meet as cluster sites. The FK cluster widths used were $W/2 = 64, 128, 256, 512, 1024, 2048$.

To have proper FK clusters we require equilibration in the SW algorithm. We numerically determined  that the equilibration time for $Q=2, 3, 4$ is of the order of $W$ spin updates by looking at the relaxation of the average energy per spin and the average largest cluster size.  For  $Q = 2,3$ and small $W$ we ran a separate simulation to  equilibrium for  each spanning cluster which was added to our ensemble.  For $Q=2$ and $3$ and $W=2048$ and $4096$ and for all of the $Q=4$ clusters, the equilibration time was too large to  proceed in this way.  In these cases we equilibrated the system once and recorded an ensemble of spanning clusters as the simulation proceeded. We conservatively estimate the correlation time as $50$ spin updates for all $W$ and $Q$. This means we recorded a spanning cluster every $50$ spin updates. 

For each system size we grew a number of clusters. For all  $Q$ our ensemble was  $2000, 2000, 1000, 1000, 400, 100$ clusters for $W = 128, 256, 512, 1024, 2048, 4096$, respectively.

\section{Measuring small probabilities with random walkers}

\subsection{Previous Methods}
Small probabilities in the harmonic measure correspond to very unlikely paths. As the simulation proceeds we can think of the event of a random walker landing where the measure is very small as a rare event. Thus computing small probabilities is a similar task to finding the rate of a rare chemical reaction \cite{vanKampen01}, a rare extinction of a disease \cite{Andersson00} or a population \cite{Bartlett60}, or the failure of a queuing system via queue over-flow \cite{Medhi03}.

 Accelerated numerical methods for these problems often involve biased event sampling. The sampling can frequently be cast as a random walk, either through state space or in our case, physical space. For example, one could ask what is the probability that a random walker starting half-way up a hill will successfully climb up to the top before sliding down to the bottom.  If the hill is steep, it could be impossible to directly sample the probability to climb the hill.  One could place barriers uniformly on the hill, which when crossed by the random walker, will split the random-walker into two walkers, each with equal weight which add up to the original weight of the walker.  This will aid sampling of the events higher up on the hill.  This method is called ``splitting" and effectively performs importance sampling \cite{Hammersley65}. One significant drawback of splitting is that if the barriers are too densely or sparsely spaced, the number of random walkers will tend to diverge or extinguish, respectively.  

The methods we detail in this paper are related to the splitting method, but differ in that our methods do not have the possibility of diverging or extinguishing.  Another popular method called ``milestoning" \cite{Farajian04}, does not have a divergence problem, but does require the system studied to be in equilibrium and the location of the barrier to be known \textit{a priori}, whereas our method works for equilibrium and non-equilibrium systems and the barriers are placed `on the fly.'

\subsection{Signposts}
We have developed several accelerated methods for the harmonic measure problem. The motivation, as we have stated, is  that  it is usually impossible to send in enough random walkers  to directly obtain the harmonic measure: the clusters will frequently have regions with probabilities of being hit that are smaller than $10^{-100}$. It would require of order $10^{100}$ random walkers to sample this region; such a computation is clearly impossible. 

We now review the first method we developed, the signpost method \cite{Adams08}.  The signpost method consists of two steps which are applied iteratively; see Figure \ref{fig:Signpost}. In the first (probe) step we release $N$ diffusing random walkers far from the cluster to determine which regions  are rarely visited in straightforward sampling. Next, we block off all poorly sampled regions with signposts (absorbing lines). In the second (measurement) step, $N$ more walkers are released far from the cluster and either absorb on the cluster (or the accessible perimeter) or onto the signposts. The walkers sent in this step have their weight permanently added to the harmonic measure of the perimeter sites where they landed. In the next probe step, the walkers are released from the points on the signposts where the walkers in the previous measurement landed. The new walkers have a weight of $p/N$, where $p$ is the fraction of random walkers that absorb onto signpost lines in the previous step, to conserve probability. The probe step again helps determine which regions are still poorly sampled, which are subsequently blocked off. Next, another measurement step is performed. This process is repeated until all regions are explored by the random walkers. This algorithm can be applied to on- and off-lattice clusters.  

\begin{figure}[t]
\includegraphics[width=0.48\textwidth]{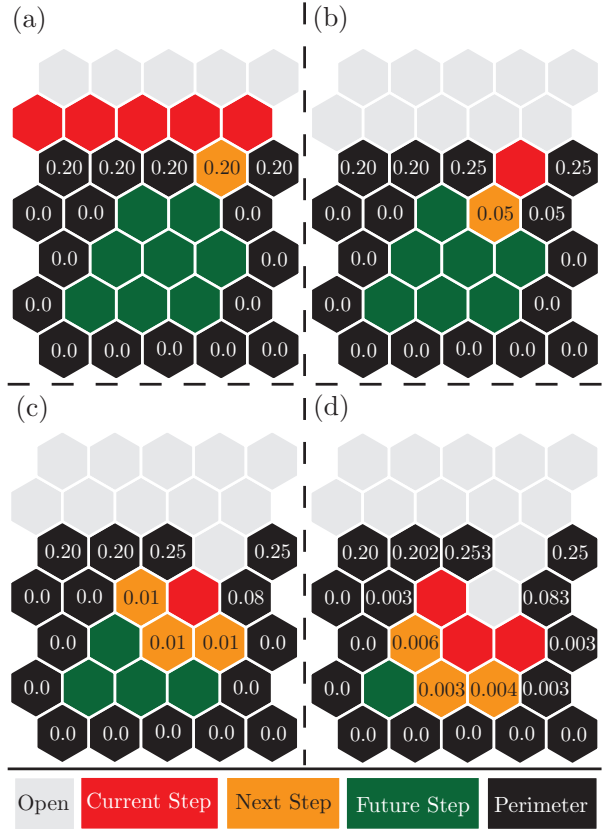}
\caption{\label{fig:Etching}  (Color online) The etching method. Walkers are released from the current level sites. The next level of soft sites absorb walkers; they are then relabeled as current level sites. Future sites are all sites which will eventually become current level sites. The first round of random walkers are launched from the row above the cluster, (a). The weight of all of the walkers released is $0.2/N$, where $N$ is the number of walkers released per current level site.  $20\%$ of the walker weight is deposited onto the top row of perimeter sites and the next level soft site, which will release $N$ walkers in the next step.  (b). One more  perimeter site is accessible to the random walkers and $5$\% of the weight is deposited on the  site in the next level. (c) Three sites in the next level each absorb $1$\% of the walker weight. (d) Due to the reduced weight of the walkers released in the next step, small probabilities are measured on the newly exposed perimeter sites.}
\end{figure}

We should note some things about this method.  First, one must determine the entire perimeter of the cluster at the beginning of the computation in order to figure out how to block poorly sampled regions.  Also, one needs to choose a rate to reduce the threshold for calling areas ``poorly sampled" in each iteration. In \cite{Adams08}, we moved the threshold down by a power of $10$ each iteration, whereas in \cite{Adams09}, we reduced it as a function of how many walkers hit the signpost in the previous iteration.  When more walkers hit the signposts we moved them even deeper.  The second method gave more consistent walker saturation, which should lead to a slower compounding of error. It is important to note the signpost algorithm is only practical for two-dimensional problems. For higher dimensions, one would need to define signpost \emph{surfaces} to block poorly sampled regions. This is could be very complex for a complicated cluster.  

\subsection{Etching}
We now describe the method we use here which we call  ``etching."  Consider the hull of FK clusters grown on a triangular lattice with periodic boundary conditions. We want to find the harmonic measure of the top perimeter from above.  To do this, we start by marking all sites that are exterior to the cluster from above as \emph{soft} sites; the soft sites are absorbing like the cluster (or accessible perimeter) sites. The highest row is limited to one level above the highest point on the perimeter. 

We next relabel every site on that highest row as a \emph{current level} site; these are not absorbing. We release $N$ random walkers, each with weight $1 / (N W)$, from each current level site. The walkers released from these sites are allowed to walk until they deposit their weight onto a soft site or a perimeter site.  If they move one level further away from the cluster, they are immediately moved back onto the current level sites using a Green's function which must be determined in advance. However, this is rather simple since it is the Green's function to return to a plane from one site above the plane.  This Green's function is used for the entire simulation and limits how far a walker can backtrack to at most one level above the cluster. After all random walkers are released, the labels on each current level site are removed and every soft site hit in the previous step is labeled as a current level site.  From each current level site $i$ we release $N$ random walkers with weight $p_i$, where $p_i$ is the amount of probability deposited on the site in the previous step divided by $N$. This process is repeated until there are no more soft sites. See Figure \ref{fig:Etching}. 

Etching can be thought of the limit of the signpost method with the signposts spaced one site apart.  However, etching has several benefits over the signpost method. First, the entire perimeter of the cluster does not need to be mapped out before we start. Both algorithms have the same time complexity, $O(W^3)$ for the complete perimeter of Ising clusters, and both methods have similar memory requirements. In contrast to the signpost method, the etching method can be easily generalized to higher-dimensional lattice problems and networks. We have successfully used etching to obtain the harmonic measure of three-dimensional percolation clusters \cite{Adams09b}.  

\begin{figure}[t]
\includegraphics[width=0.45\textwidth]{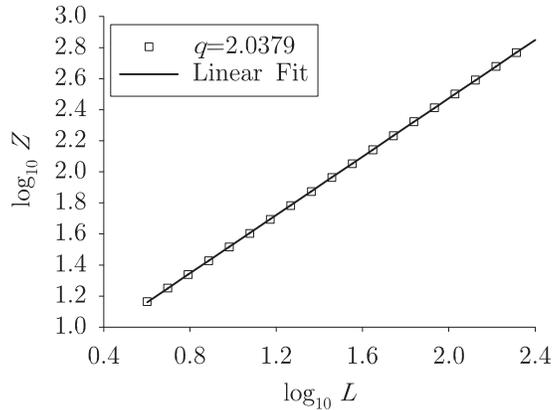}
\caption{\label{fig:Zq} An example of the fit of $\log_{10} Z(L,q)$ versus $\log_{10} L$ to a straight line for $q = 2.0379$. The behavior is similar for all $q$ values that we have examined. The slope of the line is $\tau(q) \equiv (q-1)D(q)$.}
\end{figure}
 
\subsection{Green's functions}

We have also developed a rare-event method which  may be significantly more efficient than etching and signposting for some problems. Thus far we have applied this method only to simple test problems.  This method  manipulates probabilities directly and does not allow backtracking of probability. To do this,  we calculate the Green's function $G(i,j;k,l)$, i.e., the probability to move to any of the sites $i,j$ in the next level from a given site $k,l$ in the current level. 

To illustrate our algorithm, consider finding the probability distribution a channel with absorbing walls on a square lattice. The initial condition is that the probability is uniformly distributed among the sites in the first row of  the channel and the zeroth row is a reflecting boundary.  All sites that initially have probability are denoted by  $C$. The previous level  sites, absorbing sites, and next level sites accessible to the current level sites are denoted by $B$, $A$, and $N$, respectively. (Initially, the previous level is the reflecting boundary.) In each iteration, the goal is to move all of the probability from  each current level site to the all the next level and absorbing sites. 

We find the Green's function by iteration on an index $s$. The process begins for some current level site,   $k, l$; $(k,l) \in C$.  Initially, probability only resides at $k,l$ so that for  $s=0$,  $G^s(i,j;k,l) =  \delta_{i,k} \delta_{j,l}$. In each iteration, the probability is moved to each of the current level site's neighbors
\begin{equation}
\label{eq:update}
G^{(s+1)}(i,j;k,l) = \sum_{(m,n)} W(i,j;m,n) G^{(s)}(m,n;k,l), 
\end{equation}
using the jump probability, 
\begin{eqnarray}
W(i,j;m,n) &=& \frac{1}{4} ( \delta_{i,m+1}\delta_{j,n} + \delta_{i,m-1}\delta_{j,n} + \delta_{i,m}\delta_{j,n+1}  \nonumber \\
 &+&  \delta_{i,m}\delta_{j,n-1} )  \quad (m,n) \in C \nonumber \\
  &=& G_B(i,j;m,n) \quad (m,n) \in B \nonumber \\
  &=& \delta_{i,m}\delta_{j,n} \quad (m,n) \in A \cup N
\end{eqnarray}
Here $G_B(i,j;m,n)$ is the Green's function for the previous level, see below, and the last line represents the probability staying at absorbing and next level sites. $G_B(i,j;m,n)$ takes into account all the processes that would correspond to random walkers backtracking before the previous level. To start the process, the reflecting boundary has $G_B(i,j;0,n) = \delta_{i,1}\delta_{j,n}$.

\begin{figure*}
\includegraphics[width=0.85\textwidth]{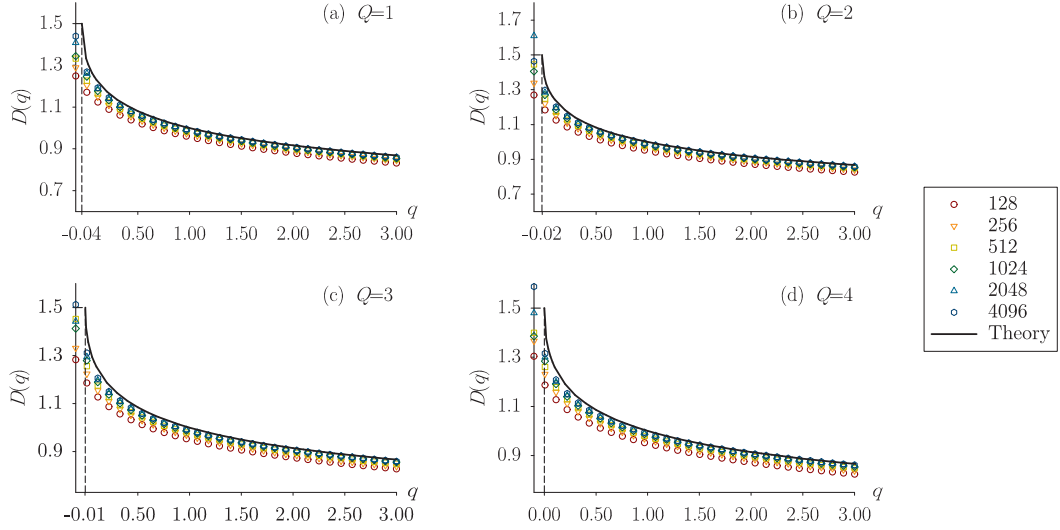}
\caption{\label{fig:D_q_Q1234}  (Color online) The $D(q)$ spectrum for the accessible perimeters of $Q=1,2,3,4$ clusters, in (a), (b), (c), and (d), respectively. The solid lines are the theory of  \cite{Duplantier00} and the symbols are the results of our simulations for several system widths. The vertical dotted lines marks $q_{min}$  for the theoretical spectra for infinite systems.}
\end{figure*}

For large $s$, virtually all of the probability will be on absorbing sites and next level sites. In any finite amount of time, some slight probability will remain in the current level, so after some stopping criteria is met, the probabilities recorded on the absorbing and next level sites must be normalized. When this has been achieved, we have the Green's function from a given site in the current level, $k,l$, to any site in the next level, $i,j$:
\begin{equation}
\label{eq:GreenNext}
G_B(i,j;k,l) = \lim_{s \rightarrow \infty } G^{(s)}(i,j;k,l).
\end{equation}
In the next step, this $G_B$ will be used as a jump probability.  

This process is repeated for all current level sites so that Green's functions from those sites to the next level sites and absorbing sites are calculated. With these Green's functions, it is easy to determine where the probability from the first level will end up. If the probability in the starting level is $P(k,l)$, then the probability in the next level is:
\begin{equation}
\label{eq:ProbAbsorbing}
P(i,j) = \sum_{ (k,l) \in C } G_B(i,j;k,l) P(k,l)
\end{equation}
Note that  $(i,j)$ can be  absorbing sites as well as  next level sites.

The next step is to relabel all current level sites as previous level sites,  relabel all next level sites as current level sites, and mark all sites that are accessible to the new current level sites (which are not previous or absorbing sites) as next level sites. Then the process is repeated.
 
%This process is summarized in the following procedure.
%\begin{enumerate}

%\item For each site in the current level, start with all the probability at that site: $G^0(i,j;k,l) =\delta_{i,k} \delta_{j,l}$ . Repeatly use (\ref{eq:update}) to move the probability until practically no probablity is left on current level sites. Next, use (\ref{eq:GreenNext}) to calculate the new Green's functions for each of the current level sites to every next level and absorbing site. 

%\item Use (\ref{eq:ProbAbsorbing}) to determine how the initial will be distributed among the next level and absorbing sites. 

%\item Relabel the current and next level sites,  $[(i,j) \in C] \rightarrow [(i,j) \in B]$ and $[(i,j) \in C] \rightarrow [(i,j) \in B]$. Also determine the new next level sites, if any exist. If there are next level sites repeat step 1.

%\end{enumerate}

The end result of this process is that all of the original probability is at absorbing sites, as it would be using signposting or etching. Although this example contained only sites that were completely absorbing or non-absorbing, the Green's function method can easily be generalized to partial absorption problems. 

The Green's function method is somewhat more complex to program than the etching method and  the simplest implementation involves setting up the Green's function look-ups in sparse arrays. This leads to a memory complexity which grows like $W^{2d}$, where $d$ is the dimension of the space. The memory complexity would significantly reduce its usefulness, as it would take at least one terabyte to store a two-dimensional cluster with a length scale of $1000$ lattice sites.  However, it is possible to store the Green's function lookup in an associative array; this reduces the memory complexity to $W^{d-1 + D}$, where $D$ is the fractal dimension of the perimeter. For the external perimeter of two-dimensional percolation clusters the memory complexity grows like $W^{7/3}$, which is quite close to the memory complexity for etching, $W^2$.  For a cluster with a length scale of $1000$ sites, the minimum required memory would be about ten megabytes for the Green's function method. 

\begin{figure}
\includegraphics[width=0.48\textwidth]{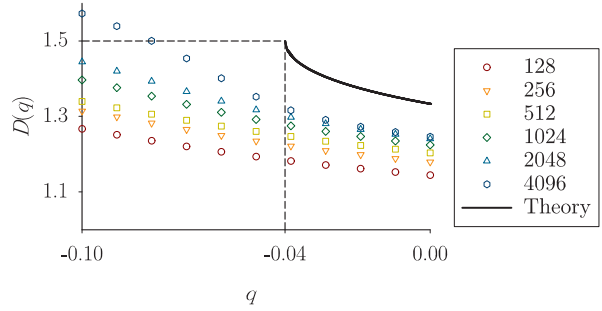}
\caption{\label{fig:D_q_Q1_fine}  (Color online) The $D(q)$ spectrum for the accessible perimeters of $Q=1$ clusters for small $q$. As the system size increases the simulated values increase, presumably to approach infinity for $q< -1/24$.}
\end{figure}

\begin{figure}

\includegraphics[width=0.4\textwidth]{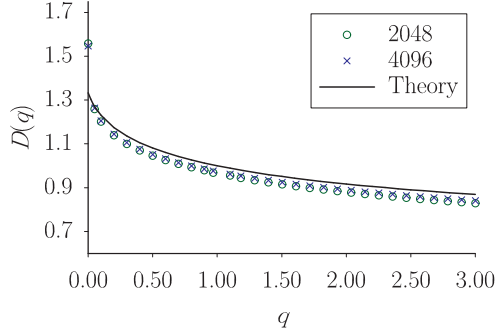}
\caption{\label{fig:D_q_Q1_complete}  (Color online) The $D(q)$ spectrum for $Q=1$ for the complete perimeter. There is no theoretical prediction for this quantity. However, for $q$ substantially bigger than 0 we expect this result to be very similar to the result for the accessible perimeter since large probabilities will dominate the sum in Eq. (\ref{eq:Zq}). The line labeled ``theory" is for the accessible perimeter.}
\end{figure}

\section{Results}
We used etching  to find the harmonic measure of $Q$-state Potts model clusters. We analyze the measure by producing $D(q)$ spectra and histograms of the probability distributions. To obtain $D(q)$, we start by sectioning individual clusters into boxes of length $L$ as described above. Because we are using a triangular lattice, it is convenient to use a parallelogram aligned with the lattice as a box. After completely tiling the cluster with boxes, we define the probability within a box $p_{i,L}$ as the sum of the measure of   perimeter sites within the box. We then calculate $Z(L,q)$ using Eq.~(\ref{eq:Zq}).  $D(q)$ is related to  $Z(L,q)$ by $(q-1)D(q) = m$, where $m$ is the slope of $\log Z(L,q)$ versus $\log L$. 

We found that for a given $Q$ and $q$, all system sizes have similar local slope behavior over a range of $L$; see Figure \ref{fig:Zq}. In order to average over the ensemble we average $\log Z$. However, if we use the slopes for each individual member of the ensemble and average them we get virtually identical results. 

The spectra of generalized dimensions for the external hulls of $Q=1-4$ are given in Figure \ref{fig:D_q_Q1234}. In all cases the results are close to the theoretical predictions \cite{Duplantier00}. The theoretical predictions include a divergence of $D(q)$ for $q < q_{min}$ for an infinite system, see below. Our simulation results increase rapidly with $W$ for this regime, as expected; see Figure \ref{fig:D_q_Q1_fine}. 

For completeness, we include the spectrum of generalized dimensions for the complete perimeter for the case $Q=1$; see Figure~\ref{fig:D_q_Q1_complete}. There is no theoretical prediction for this quantity. For positive $q$ the results are close to those of the accessible perimeter shown in Figure \ref{fig:D_q_Q1234}.  This is because, for positive $q$, large probabilities contribute most of the weight in $Z(q)$. Near $q=0$ the two spectra differ because there are significantly more sites with small measure for the complete hulls.

\begin{figure}
\includegraphics[width=0.45\textwidth]{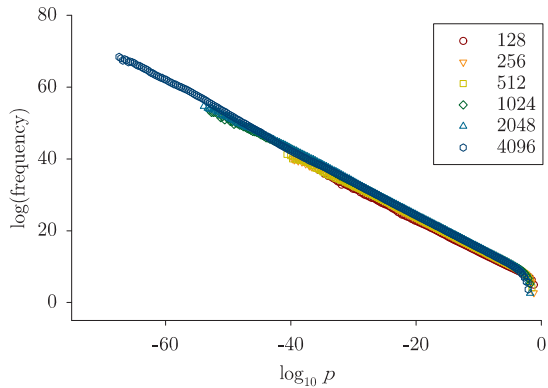}
\caption{\label{fig:rawhist}  (Color online) The histogram of the frequency of occurrence  of the values of $p$ for the accessible perimeter for $Q=1$. The points for various values of $W$ are superimposed.}
\end{figure}

\begin{figure*}
\includegraphics[width=0.9\textwidth]{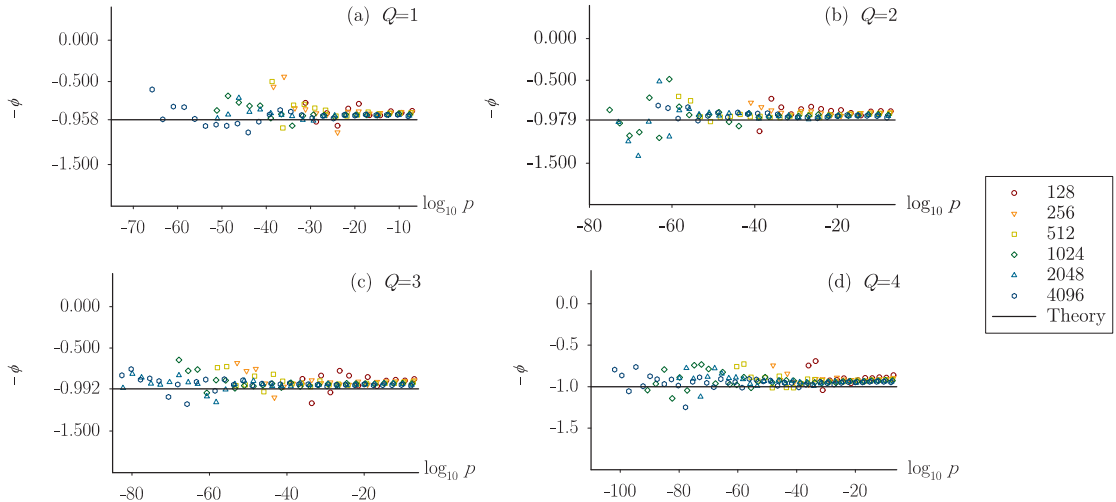}
\caption{\label{fig:ProbDistQ1234}  (Color online) The local slope of the histogram of the frequency of occurrence  of the values of $p$ for the accessible perimeter for $Q=1,2,3,4$ in (a), (b), (c), and (d), respectively. Also shown (solid lines) are the theoretical predictions of the local slope from \cite{Duplantier00}. Note that in (b) the smallest probabilities recorded were not from the largest system size, but were from $W=1024$. This can be understood by the fact that ten times as many clusters were generated for $W=1024$. That is, among the many samples at $W=1024$, a few abnormally deep clusters were recorded which happened to have the smallest probabilities.}
\end{figure*}

\begin{figure}[t]
\includegraphics[width=0.45\textwidth]{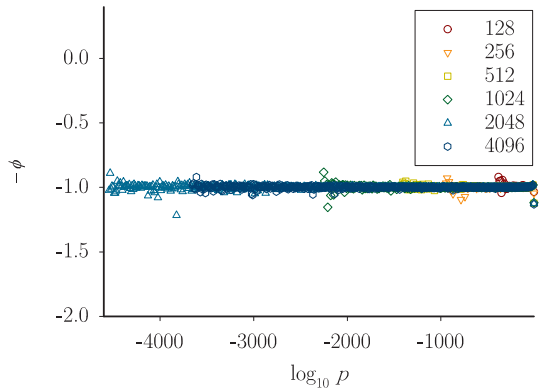}
\caption{\label{fig:ProbDistQ1complete}  (Color online) The local slope of the histogram of the frequency of occurrence  of the values of $p$ for the complete perimeter for $Q=1$.}
\end{figure}

We also considered the distribution of the values of $p$ directly, by making histograms of its frequency for all $Q$ and $W$. The histograms turn out to be power laws with negative powers near $-1$; for an example see Figure \ref{fig:rawhist}.  Since the histogram is very accurately a power-law in $p$,  it is useful to plot the local slope of the histogram, which is shown in  Figure \ref{fig:ProbDistQ1234} for the accessible perimeters for $Q=1-4$. We also show the local slope for the complete perimeter of $Q=1$; see Figure~\ref{fig:ProbDistQ1complete}. The slope is calculated over about 10 orders of magnitude in $p$ for the accessible perimeter, and more than one order of magnitude for the complete perimeter. 

The significance of the slope is that it gives information about the non-scaling aspects of the distribution, and, in particular, the value of $q_{min}$ mentioned above. If we call the slope of the histogram $-\phi$ (so that $\phi$ is a positive number) we see that the partition function of Eq. (\ref{eq:Zq}) formally diverges if $q< \phi-1$, or, said another way, we expect $D(q)$ to be undefined for $q< q_{min} = -1+\phi$. This means that the partition function is dominated by a few instances of very small probabilities which does not scale as power-law in $R/L$. 
The values  for the limit of the spectrum agree well with the predictions of Duplantier \cite{Duplantier99, Duplantier00}; see Figure \ref{fig:ProbDistQ1234}. Note that the slopes are very nearly constant over about 40 orders of magnitude in $p$.

The slopes for the complete perimeter of percolation clusters are also constant over many orders of magnitude; see Figure \ref{fig:ProbDistQ1complete}. In this case we find that $\phi$ is very close to 1, and the limit of the spectrum is at $q_{min}=0$. There is no theory for this case, and no explanation for this intriguing result.

%\begin{figure}
%\includegraphics[width=0.45\textwidth]{Histogram_External_Qone_fig12}
%\caption{\label{fig:ProbDistQ1} The local slope of the histogram of the frequency of occurrence  of the values of $p$ for the accessible perimeter for $Q=1$. Also shown (solid line) is the theoretical prediction of the local slope from \cite{Duplantier00}.}
%\end{figure}

%\begin{figure}[b]
%\includegraphics[width=0.45\textwidth]{Histogram_External_Qtwo_fig13}
%\caption{\label{fig:ProbDistQ2} The local slope of the histogram of the frequency of occurrence  of the values of $p$ for the accessible perimeter for $Q=2$. Also shown (solid line) is the theoretical prediction of the local slope from \cite{Duplantier00}.}
%\end{figure}

%\begin{figure}
%\includegraphics[width=0.45\textwidth]{Histogram_External_Qthree_fig14}
%\caption{\label{fig:ProbDistQ3} The local slope of the histogram of the frequency of occurrence  of the values of $p$ for the accessible perimeter for $Q=3$. Also shown (solid line) is the theoretical prediction of the local slope from \cite{Duplantier00}.}
%\end{figure}

%\begin{figure}
%\includegraphics[width=0.45\textwidth]{Histogram_External_Qfour_fig15}
%\caption{\label{fig:ProbDistQ4} The local slope of the histogram of the frequency of occurrence  of the values of $p$ for the accessible perimeter for $Q=4$. Also shown (solid line) is the theoretical prediction of the local slope from \cite{Duplantier00}.}
%\end{figure}

\section{Error estimate}
Since etching involves sampling the probability, there will be errors due to the finite number of random walkers released at each step. 
For the results in this paper, we released $10^3$ random-walkers per current level site for all system widths and $Q$ values. 

We can estimate the sampling errors as follows: we considered one percolation cluster with $W=2048$ and made 10 independent computations of the $p_i$. The variance of the probability over this sample at a given point on the cluster, $\delta p_i$,  is a measure of the reliability of the measurement. In our case we found that some points have a rather large percentage error, though always less than a  factor of 3, but the average over all the points, $\langle \delta p_i/p_i\rangle$, was 23\%. Note that the very small probabilities well inside the cluster have very small errors. There is no build-up of the error as we etch toward the interior, as might have been expected. 
 
 If it is necessary to reduce the error further, more random walkers can be used. However, we believe that the ensemble averaging that we did means that the generalized dimensions are much more accurate than the individual probabilities. Our evidence for the last statement is the good quality of the fit in Figure \ref{fig:Zq}, and the closeness of the results in Figure \ref{fig:D_q_Q1234} to theory. Note also that $D(0)$ is close to the known fractal dimensions of the exterior perimeters.

\section{Conclusions}
In this paper, we presented the etching method,  a new accelerated technique for computing the harmonic measure.  We are able to measure probabilities as small as $10^{-4600}$. We showed how this method relates to other methods. We used etching to obtain the harmonic measure for the accessible perimeter of FK clusters for the  $Q$-state Potts model for $Q=1-4$, for a range of system sizes. We compared this data to theoretical predictions \cite{Duplantier99,Duplantier00}. These theories were produced for a continuum model which, in principle, might not apply to the scaling limit of the $Q$-state Potts model on a lattice. In fact, we found good agreement between our numerical results and the theoretical predictions for every comparison we made, including the $D(q)$ spectra and the slopes of the power-law probability distributions. 

For the complete perimeter of percolation clusters, we found the slope to be almost exactly $-1$ for about 4000 orders of magnitude. This suggests the smallest $q$ for which $D(q)$ is defined is $q=0$. This means that there are many instances of  small probabilities on the complete perimeter of percolation clusters which tend to diverge towards negative infinity faster than any power of $R/L$. 

Etching, signposting, and the Green's function method are three tools which can find very small probabilities. The advantage of signposting is that it is natural to use in off-lattice systems, and, in fact, we have applied it to off-lattice DLA \cite{Adams09}. Etching is simple to program and should easy to use in higher dimensional on-lattice systems. Lastly, the Green's function method is likely to be the most efficient of the algorithms for on-lattice and network systems, but it is more difficult to implement and requires more memory than etching. The etching and Green's function methods (but not signposting) can be used in problems which involve absorption probabilities less than unity.

\section{Acknowledgements}

This work was supported in part by  National Science Foundation grant DMS-0553487.

\bibliography{Etching_Draft_4}% Produces the bibliography via BibTeX.

\begin{thebibliography}{37}
\expandafter\ifx\csname natexlab\endcsname\relax\def\natexlab#1{#1}\fi
\expandafter\ifx\csname bibnamefont\endcsname\relax
  \def\bibnamefont#1{#1}\fi
\expandafter\ifx\csname bibfnamefont\endcsname\relax
  \def\bibfnamefont#1{#1}\fi
\expandafter\ifx\csname citenamefont\endcsname\relax
  \def\citenamefont#1{#1}\fi
\expandafter\ifx\csname url\endcsname\relax
  \def\url#1{\texttt{#1}}\fi
\expandafter\ifx\csname urlprefix\endcsname\relax\def\urlprefix{URL }\fi
\providecommand{\bibinfo}[2]{#2}
\providecommand{\eprint}[2][]{\url{#2}}

\bibitem[{\citenamefont{Wu}(1982)}]{Wu82}
\bibinfo{author}{\bibfnamefont{F.~Y.} \bibnamefont{Wu}}, \bibinfo{journal}{Rev.
  Mod. Phys.} \textbf{\bibinfo{volume}{54}}, \bibinfo{pages}{235}
  (\bibinfo{year}{1982}).

\bibitem[{\citenamefont{Stauffer and Aharony}(1994)}]{StaufferBook94}
\bibinfo{author}{\bibfnamefont{D.}~\bibnamefont{Stauffer}} \bibnamefont{and}
  \bibinfo{author}{\bibfnamefont{A.}~\bibnamefont{Aharony}},
  \emph{\bibinfo{title}{Introduction to Percolation Theory}}
  (\bibinfo{publisher}{CRC press}, \bibinfo{year}{1994}).

\bibitem[{\citenamefont{Fortuin and Kasteleyn}(1972)}]{Fortuin72}
\bibinfo{author}{\bibfnamefont{C.~M.} \bibnamefont{Fortuin}} \bibnamefont{and}
  \bibinfo{author}{\bibfnamefont{P.~W.} \bibnamefont{Kasteleyn}},
  \bibinfo{journal}{Physica} \textbf{\bibinfo{volume}{57}},
  \bibinfo{pages}{536} (\bibinfo{year}{1972}).

\bibitem[{\citenamefont{Kasteleyn and Fortuin}(1969)}]{Kasteleyn69}
\bibinfo{author}{\bibfnamefont{P.~W.} \bibnamefont{Kasteleyn}}
  \bibnamefont{and} \bibinfo{author}{\bibfnamefont{C.~M.}
  \bibnamefont{Fortuin}}, \bibinfo{journal}{J. Phys. Soc. of Japan}
  \textbf{\bibinfo{volume}{26}}, \bibinfo{pages}{11} (\bibinfo{year}{1969}).

\bibitem[{\citenamefont{Mandelbrot}(1982)}]{Mandelbrot82}
\bibinfo{author}{\bibfnamefont{B.~B.} \bibnamefont{Mandelbrot}},
  \emph{\bibinfo{title}{{The Fractal Geometry of Nature}}}
  (\bibinfo{publisher}{W. H. Freeman}, \bibinfo{year}{1982}).

\bibitem[{\citenamefont{Hasley and Leibig}(1992)}]{Halsey92}
\bibinfo{author}{\bibfnamefont{T.~C.} \bibnamefont{Hasley}} \bibnamefont{and}
  \bibinfo{author}{\bibfnamefont{M.}~\bibnamefont{Leibig}},
  \bibinfo{journal}{Ann. Phys.} \textbf{\bibinfo{volume}{219}},
  \bibinfo{pages}{109} (\bibinfo{year}{1992}).

\bibitem[{\citenamefont{Sapoval et~al.}(2001)\citenamefont{Sapoval, Andrade,
  and Filoche}}]{Sapoval01}
\bibinfo{author}{\bibfnamefont{B.}~\bibnamefont{Sapoval}},
  \bibinfo{author}{\bibfnamefont{J.~S.} \bibnamefont{Andrade}},
  \bibnamefont{and} \bibinfo{author}{\bibfnamefont{M.}~\bibnamefont{Filoche}},
  \bibinfo{journal}{Chem. Eng. Sci.} \textbf{\bibinfo{volume}{56}},
  \bibinfo{pages}{5011} (\bibinfo{year}{2001}).

\bibitem[{\citenamefont{Witten and Sander}(1981)}]{Witten81}
\bibinfo{author}{\bibfnamefont{T.~A.} \bibnamefont{Witten}} \bibnamefont{and}
  \bibinfo{author}{\bibfnamefont{L.~M.} \bibnamefont{Sander}},
  \bibinfo{journal}{Phys. Rev. Lett.} \textbf{\bibinfo{volume}{47}},
  \bibinfo{pages}{1400} (\bibinfo{year}{1981}).

\bibitem[{\citenamefont{Mandelbrot and Evertsz}(1990)}]{Mandelbrot90}
\bibinfo{author}{\bibfnamefont{B.~B.} \bibnamefont{Mandelbrot}}
  \bibnamefont{and} \bibinfo{author}{\bibfnamefont{C.~J.~G.}
  \bibnamefont{Evertsz}}, \bibinfo{journal}{Nature}
  \textbf{\bibinfo{volume}{348}}, \bibinfo{pages}{143} (\bibinfo{year}{1990}).

\bibitem[{\citenamefont{Duplantier}(1999)}]{Duplantier99}
\bibinfo{author}{\bibfnamefont{B.}~\bibnamefont{Duplantier}},
  \bibinfo{journal}{Phys. Rev. Lett.} \textbf{\bibinfo{volume}{82}},
  \bibinfo{pages}{3940} (\bibinfo{year}{1999}).

\bibitem[{\citenamefont{Duplantier}(2000)}]{Duplantier00}
\bibinfo{author}{\bibfnamefont{B.}~\bibnamefont{Duplantier}},
  \bibinfo{journal}{Phys. Rev. Lett.} \textbf{\bibinfo{volume}{84}},
  \bibinfo{pages}{1363} (\bibinfo{year}{2000}).

\bibitem[{\citenamefont{Belikov et~al.}(2008)\citenamefont{Belikov, Gruzberg,
  and Rushkin}}]{Belikov08}
\bibinfo{author}{\bibfnamefont{A.}~\bibnamefont{Belikov}},
  \bibinfo{author}{\bibfnamefont{I.~A.} \bibnamefont{Gruzberg}},
  \bibnamefont{and} \bibinfo{author}{\bibfnamefont{I.}~\bibnamefont{Rushkin}},
  \bibinfo{journal}{J. Phys. A} \textbf{\bibinfo{volume}{41}},
  \bibinfo{pages}{285006} (\bibinfo{year}{2008}).

\bibitem[{\citenamefont{Duplantier and Binder}(2008)}]{Duplantier08}
\bibinfo{author}{\bibfnamefont{B.}~\bibnamefont{Duplantier}} \bibnamefont{and}
  \bibinfo{author}{\bibfnamefont{I.~A.} \bibnamefont{Binder}},
  \bibinfo{journal}{Nucl. Phys. B} \textbf{\bibinfo{volume}{802}},
  \bibinfo{pages}{494} (\bibinfo{year}{2008}).

\bibitem[{\citenamefont{Bettelheim et~al.}(2005)\citenamefont{Bettelheim,
  Rushkin, Gruzberg, and Wiegmann}}]{Bettelhiem05}
\bibinfo{author}{\bibfnamefont{E.}~\bibnamefont{Bettelheim}},
  \bibinfo{author}{\bibfnamefont{I.}~\bibnamefont{Rushkin}},
  \bibinfo{author}{\bibfnamefont{I.~A.} \bibnamefont{Gruzberg}},
  \bibnamefont{and} \bibinfo{author}{\bibfnamefont{P.}~\bibnamefont{Wiegmann}},
  \bibinfo{journal}{Phys. Rev. Lett.} \textbf{\bibinfo{volume}{95}},
  \bibinfo{pages}{170602} (\bibinfo{year}{2005}).

\bibitem[{\citenamefont{Gruzberg}(2006)}]{Gruzberg06}
\bibinfo{author}{\bibfnamefont{I.~A.} \bibnamefont{Gruzberg}},
  \bibinfo{journal}{J. Phys. A} \textbf{\bibinfo{volume}{39}},
  \bibinfo{pages}{12601} (\bibinfo{year}{2006}).

\bibitem[{\citenamefont{Meakin et~al.}(1986)\citenamefont{Meakin, Coniglio,
  Stanley, and Witten}}]{Meakin86}
\bibinfo{author}{\bibfnamefont{P.}~\bibnamefont{Meakin}},
  \bibinfo{author}{\bibfnamefont{A.}~\bibnamefont{Coniglio}},
  \bibinfo{author}{\bibfnamefont{H.~E.} \bibnamefont{Stanley}},
  \bibnamefont{and} \bibinfo{author}{\bibfnamefont{T.~A.}
  \bibnamefont{Witten}}, \bibinfo{journal}{Phys. Rev. A}
  \textbf{\bibinfo{volume}{34}}, \bibinfo{pages}{3325} (\bibinfo{year}{1986}).

\bibitem[{\citenamefont{Ball and Spivack}(1990)}]{Ball90}
\bibinfo{author}{\bibfnamefont{R.~C.} \bibnamefont{Ball}} \bibnamefont{and}
  \bibinfo{author}{\bibfnamefont{O.~R.} \bibnamefont{Spivack}},
  \bibinfo{journal}{J. Phys. A} \textbf{\bibinfo{volume}{23}},
  \bibinfo{pages}{5295} (\bibinfo{year}{1990}).

\bibitem[{\citenamefont{Hanan and Heffernan}(2008)}]{Hanan08}
\bibinfo{author}{\bibfnamefont{W.~G.} \bibnamefont{Hanan}} \bibnamefont{and}
  \bibinfo{author}{\bibfnamefont{D.~M.} \bibnamefont{Heffernan}},
  \bibinfo{journal}{Phys. Rev. E} \textbf{\bibinfo{volume}{77}},
  \bibinfo{pages}{011405} (\bibinfo{year}{2008}).

\bibitem[{\citenamefont{Hastings and Levitov}(1998)}]{Hastings98}
\bibinfo{author}{\bibfnamefont{M.~B.} \bibnamefont{Hastings}} \bibnamefont{and}
  \bibinfo{author}{\bibfnamefont{L.~S.} \bibnamefont{Levitov}},
  \bibinfo{journal}{Physica D} \textbf{\bibinfo{volume}{116}},
  \bibinfo{pages}{224} (\bibinfo{year}{1998}).

\bibitem[{\citenamefont{Davidovitch et~al.}(2000)\citenamefont{Davidovitch,
  Levermann, and Procaccia}}]{Davidovitch00}
\bibinfo{author}{\bibfnamefont{B.}~\bibnamefont{Davidovitch}},
  \bibinfo{author}{\bibfnamefont{A.}~\bibnamefont{Levermann}},
  \bibnamefont{and}
  \bibinfo{author}{\bibfnamefont{I.}~\bibnamefont{Procaccia}},
  \bibinfo{journal}{Phys. Rev. E} \textbf{\bibinfo{volume}{62}},
  \bibinfo{pages}{R5919} (\bibinfo{year}{2000}).

\bibitem[{\citenamefont{Davidovitch et~al.}(2001)\citenamefont{Davidovitch,
  Jensen, Levermann, Mathiesen, and Procaccia}}]{Davidovitch01}
\bibinfo{author}{\bibfnamefont{B.}~\bibnamefont{Davidovitch}},
  \bibinfo{author}{\bibfnamefont{M.~H.} \bibnamefont{Jensen}},
  \bibinfo{author}{\bibfnamefont{A.}~\bibnamefont{Levermann}},
  \bibinfo{author}{\bibfnamefont{J.}~\bibnamefont{Mathiesen}},
  \bibnamefont{and}
  \bibinfo{author}{\bibfnamefont{I.}~\bibnamefont{Procaccia}},
  \bibinfo{journal}{Phys. Rev. Lett.} \textbf{\bibinfo{volume}{87}},
  \bibinfo{pages}{164101} (\bibinfo{year}{2001}).

\bibitem[{\citenamefont{Jensen et~al.}(2002)\citenamefont{Jensen, Levermann,
  Mathiesen, and Procaccia}}]{Jensen02}
\bibinfo{author}{\bibfnamefont{M.~H.} \bibnamefont{Jensen}},
  \bibinfo{author}{\bibfnamefont{A.}~\bibnamefont{Levermann}},
  \bibinfo{author}{\bibfnamefont{J.}~\bibnamefont{Mathiesen}},
  \bibnamefont{and}
  \bibinfo{author}{\bibfnamefont{I.}~\bibnamefont{Procaccia}},
  \bibinfo{journal}{Phys. Rev. E} \textbf{\bibinfo{volume}{65}},
  \bibinfo{pages}{046109} (\bibinfo{year}{2002}).

\bibitem[{\citenamefont{Adams et~al.}(2008)\citenamefont{Adams, Sander, and
  Ziff}}]{Adams08}
\bibinfo{author}{\bibfnamefont{D.~A.} \bibnamefont{Adams}},
  \bibinfo{author}{\bibfnamefont{L.~M.} \bibnamefont{Sander}},
  \bibnamefont{and} \bibinfo{author}{\bibfnamefont{R.~M.} \bibnamefont{Ziff}},
  \bibinfo{journal}{Phys. Rev. Lett.} \textbf{\bibinfo{volume}{101}},
  \bibinfo{pages}{144102} (\bibinfo{year}{2008}).

\bibitem[{\citenamefont{Halsey et~al.}(1986)\citenamefont{Halsey, Jensen,
  Kadanoff, Procaccia, and Shraiman}}]{Halsey86}
\bibinfo{author}{\bibfnamefont{T.~C.} \bibnamefont{Halsey}},
  \bibinfo{author}{\bibfnamefont{M.~H.} \bibnamefont{Jensen}},
  \bibinfo{author}{\bibfnamefont{L.~P.} \bibnamefont{Kadanoff}},
  \bibinfo{author}{\bibfnamefont{I.}~\bibnamefont{Procaccia}},
  \bibnamefont{and} \bibinfo{author}{\bibfnamefont{B.~I.}
  \bibnamefont{Shraiman}}, \bibinfo{journal}{Phys. Rev. A}
  \textbf{\bibinfo{volume}{33}}, \bibinfo{pages}{1141 } (\bibinfo{year}{1986}).

\bibitem[{\citenamefont{Makarov}(1985)}]{Makarov85}
\bibinfo{author}{\bibfnamefont{N.}~\bibnamefont{Makarov}},
  \bibinfo{journal}{Proc. London Math. Soc.} \textbf{\bibinfo{volume}{51}},
  \bibinfo{pages}{369} (\bibinfo{year}{1985}).

\bibitem[{\citenamefont{Leath}(1976)}]{Leath76}
\bibinfo{author}{\bibfnamefont{P.~L.} \bibnamefont{Leath}},
  \bibinfo{journal}{Phys. Rev. B} \textbf{\bibinfo{volume}{14}},
  \bibinfo{pages}{5046} (\bibinfo{year}{1976}).

\bibitem[{\citenamefont{Grossman and Aharony}(1987)}]{Grossman87}
\bibinfo{author}{\bibfnamefont{T.}~\bibnamefont{Grossman}} \bibnamefont{and}
  \bibinfo{author}{\bibfnamefont{A.}~\bibnamefont{Aharony}},
  \bibinfo{journal}{J. Phys. A} \textbf{\bibinfo{volume}{20}},
  \bibinfo{pages}{1193} (\bibinfo{year}{1987}).

\bibitem[{\citenamefont{Swendsen and Wang}(1987)}]{Swendsen87}
\bibinfo{author}{\bibfnamefont{R.~H.} \bibnamefont{Swendsen}} \bibnamefont{and}
  \bibinfo{author}{\bibfnamefont{J.~S.} \bibnamefont{Wang}},
  \bibinfo{journal}{Phys. Rev. Lett.} \textbf{\bibinfo{volume}{58}},
  \bibinfo{pages}{86} (\bibinfo{year}{1987}).

\bibitem[{\citenamefont{Kim and Joseph}(1974)}]{kim1974exact}
\bibinfo{author}{\bibfnamefont{D.}~\bibnamefont{Kim}} \bibnamefont{and}
  \bibinfo{author}{\bibfnamefont{R.}~\bibnamefont{Joseph}},
  \bibinfo{journal}{J. Phys. C} \textbf{\bibinfo{volume}{7}},
  \bibinfo{pages}{L167} (\bibinfo{year}{1974}).

\bibitem[{\citenamefont{van Kampen}(2001)}]{vanKampen01}
\bibinfo{author}{\bibfnamefont{N.~G.} \bibnamefont{van Kampen}},
  \emph{\bibinfo{title}{{Stochastic Processes in Physics and Chemistry}}}
  (\bibinfo{publisher}{North-Holland}, \bibinfo{year}{2001}).

\bibitem[{\citenamefont{Andersson and Britton}(2000)}]{Andersson00}
\bibinfo{author}{\bibfnamefont{H.}~\bibnamefont{Andersson}} \bibnamefont{and}
  \bibinfo{author}{\bibfnamefont{T.}~\bibnamefont{Britton}},
  \emph{\bibinfo{title}{{Stochastic Epidemic Models and Their Statistical
  Analysis}}} (\bibinfo{publisher}{Springer}, \bibinfo{year}{2000}).

\bibitem[{\citenamefont{Bartlett}(1960)}]{Bartlett60}
\bibinfo{author}{\bibfnamefont{M.~S.} \bibnamefont{Bartlett}},
  \bibinfo{journal}{Methuen, London}  (\bibinfo{year}{1960}).

\bibitem[{\citenamefont{Medhi and Medhi}(2003)}]{Medhi03}
\bibinfo{author}{\bibfnamefont{J.}~\bibnamefont{Medhi}} \bibnamefont{and}
  \bibinfo{author}{\bibfnamefont{J.}~\bibnamefont{Medhi}},
  \emph{\bibinfo{title}{{Stochastic Models in Queueing Theory}}}
  (\bibinfo{publisher}{Academic Press Boston}, \bibinfo{year}{2003}).

\bibitem[{\citenamefont{Hammersley and Handscomb}(1965)}]{Hammersley65}
\bibinfo{author}{\bibfnamefont{J.~M.} \bibnamefont{Hammersley}}
  \bibnamefont{and} \bibinfo{author}{\bibfnamefont{D.~C.}
  \bibnamefont{Handscomb}}, \emph{\bibinfo{title}{Monte Carlo Methods}}
  (\bibinfo{publisher}{Methuen}, \bibinfo{address}{London},
  \bibinfo{year}{1965}).

\bibitem[{\citenamefont{Faradjian and Elber}(2004)}]{Farajian04}
\bibinfo{author}{\bibfnamefont{A.~K.} \bibnamefont{Faradjian}}
  \bibnamefont{and} \bibinfo{author}{\bibfnamefont{R.}~\bibnamefont{Elber}},
  \bibinfo{journal}{J. Chem. Phys.} \textbf{\bibinfo{volume}{120}},
  \bibinfo{pages}{10880} (\bibinfo{year}{2004}).

\bibitem[{\citenamefont{Adams et~al.}(2009)\citenamefont{Adams, Sander, Somfai,
  and Ziff}}]{Adams09}
\bibinfo{author}{\bibfnamefont{D.~A.} \bibnamefont{Adams}},
  \bibinfo{author}{\bibfnamefont{L.~M.} \bibnamefont{Sander}},
  \bibinfo{author}{\bibfnamefont{E.}~\bibnamefont{Somfai}}, \bibnamefont{and}
  \bibinfo{author}{\bibfnamefont{R.~M.} \bibnamefont{Ziff}},
  \bibinfo{journal}{Europhys. Lett.} \textbf{\bibinfo{volume}{3}},
  \bibinfo{pages}{20001} (\bibinfo{year}{2009}).

\bibitem[{\citenamefont{Adams et~al.}(to be published)\citenamefont{Adams,
  Sander, and Ziff}}]{Adams09b}
\bibinfo{author}{\bibfnamefont{D.~A.} \bibnamefont{Adams}},
  \bibinfo{author}{\bibfnamefont{L.~M.} \bibnamefont{Sander}},
  \bibnamefont{and} \bibinfo{author}{\bibfnamefont{R.~M.} \bibnamefont{Ziff}}
  (\bibinfo{year}{to be published}).

\end{thebibliography}

\end{document}